\documentclass[a4paper,11pt]{article}

\usepackage{jcappub} 

\usepackage[T1]{fontenc} 
\usepackage{amsmath}

\usepackage{graphicx}
\usepackage{amsmath}
\usepackage{xcolor}
\usepackage[normalem]{ulem}
\usepackage{enumitem}

\usepackage{graphicx}
\usepackage{dcolumn}
\usepackage{bm}
\usepackage{amsmath}
\usepackage{amsfonts} 
\usepackage{latexsym}
\usepackage{bbm}
\usepackage{color}
\usepackage{amssymb}
\usepackage{amsthm}
\usepackage{epsf}
\usepackage{epsfig}
\usepackage{caption3}
\usepackage{appendix}
\usepackage{cases}
\usepackage{subfigure}
\usepackage{multirow}
\usepackage{float}
\usepackage{array}

\title{Exploring Ultralight Scalar Assistance in Sterile Neutrino Dark Matter: Cold Spectrum and Unusual X/Gamma-ray Signatures}
\author[a]{Yuxuan He}
\author[a,b]{Jia Liu\footnote{Corresponding author}}
\author[a]{Xiaolin Ma}
\author[c,d]{Xiao-Ping Wang\footnote{Corresponding author} }

\affiliation[a]{School of Physics and State Key Laboratory of Nuclear Physics and Technology, Peking University, Beijing 100871, China}
\affiliation[b]{Center for High Energy Physics, Peking University, Beijing 100871, China}
\affiliation[c]{School of Physics, Beihang University, Beijing 100083, China}
\affiliation[d]{Beijing Key Laboratory of Advanced Nuclear Materials and Physics, Beihang University, Beijing 100191, China}

\emailAdd{heyx25@pku.edu.cn}
\emailAdd{jialiu@pku.edu.cn}
\emailAdd{themapku@stu.pku.edu.cn}
\emailAdd{hcwangxiaoping@buaa.edu.cn}

\abstract{
We present a scalar-driven sterile neutrino production model where the interaction with the ultralight scalar field modifies the oscillation production of sterile neutrinos in the early universe. The model effectively suppresses the production of sterile neutrinos at low temperatures due to the heavy scalar mass, resulting in a colder matter power spectrum that avoids constraints from small-scale structure observations. In this model, the dominant dark matter relic is from sterile neutrinos, with only a small fraction originating from the ultralight scalar. Furthermore, the model predicts a detectable X/$\gamma$-ray flux proportional to the cubic density of local sterile neutrinos for a light scalar mass due to the light scalar coupling to sterile neutrinos. This distinguishes our model from normal decaying dark matter, which has a linear dependence on the density. In addition, the model predicts a potential low-energy monochromatic neutrino signal that can be detectable by future neutrino telescopes.
}

\begin{document}
	\maketitle
	\flushbottom
\section{Introduction}

Sterile neutrinos have long been considered as a promising candidate for dark matter (DM), which mix with active neutrinos leading to the right relic abundance in the early universe \cite{Dodelson:1993je,Abazajian:2005gj,Boyarsky:2018tvu,Kopp:2021jlk}. In the seminal framework denoted as Dodelson-Widrow scenario (DW) \cite{Dodelson:1993je}, a natural mechanism for producing sterile neutrino DM via neutrino oscillation was proposed, where the scattering between active neutrinos and Standard Model (SM) particles in the thermal bath results in the production of a suitable amount of sterile neutrinos. This same mixing angle also leads to the radioactive decay of sterile neutrinos to photons and neutrinos, which provides a  monochromatic photon signature to look for.
Despite the attractiveness of DW scenario, it has been challenged by various experiments.  The monochromatic X/$\gamma$-ray searches of distant galaxies and dwarfs set stringent limits on the mixing angle or the lifetime of the decaying sterile neutrino \cite{Boyarsky:2008ju, Horiuchi:2013noa, Bulbul:2014sua, Boyarsky:2014jta, Boyarsky:2014ska, Tamura:2014mta,Malyshev:2014xqa,Hitomi:2016mun}. Decaying sterile neutrino DM was proposed to explain the 3.5 keV line anomaly~\cite{Boyarsky:2014ska, Boyarsky:2014jta}
It is also worth noting that although the 3.5 keV line anomaly can be explained by decaying sterile neutrino DM, however this possibility was constrained by blank-sky observations and under debate~\cite{Dessert:2018qih, Abazajian:2020unr}.

Sterile neutrino DM generated by the DW mechanism or its variants is typically classified as warm dark matter (WDM). Due to free-streaming, the matter power spectrum at small scales is damped. Observations of the small-scale matter power spectrum in Lyman-$\alpha$ \cite{Viel:2005qj,Boyarsky:2009ix} and gravitational lensing \cite{Zelko:2022tgf} have placed constraints on the mass of DW sterile neutrino DM, suggesting it should be above 30 keV \cite{Yeche:2017upn} and possibly even higher \cite{Zelko:2022tgf}. On the other hand, the phase space density of Milky Way dwarf galaxies suggests a lower limit of 2.5 keV, while subhalo counts of M31 analogues give a limit of 8.8 keV \cite{Horiuchi:2013noa}. The most stringent constraint arises from a comprehensive analysis of the Lyman-$\alpha$ forest, strong gravitational lensing, and Milky Way satellites. This analysis excludes sterile neutrinos produced through the DW mechanism with a mass less than 92 keV at a 95\% confidence level\cite{Viel:2013fqw,Irsic:2017ixq,Gilman:2019nap,Nadler:2021dft,Zelko:2022tgf}.

To address the tensions and expand the parameter space for oscillation production of sterile neutrinos, several ideas have been proposed, including introducing beyond Standard Model (BSM) interactions \cite{Shi:1998km, Abazajian:2014gza, DeGouvea:2019wpf, Kelly:2020aks,Alonso-Alvarez:2021pgy,Guo:2022hnx,An:2023mkf}. These mechanisms aim to modify the effective mixing angle, allowing sterile neutrinos to be produced at an appropriate rate while evading current X/$\gamma$-ray constraints. Besides, there exist extensive explorations about the interplay of ultralight scalar and sterile neutrino~\cite{Berlin:2016bdv,Berlin:2016woy,Krnjaic:2017zlz,Brdar:2017kbt,Liao:2018byh,Ge:2018uhz,Farzan:2019yvo,Huang:2021kam,Losada:2021bxx,Huang:2022wmz,Dev:2022bae,Gan:2023wnp,Davoudiasl:2023uiq,Bezrukov:2018wvd,Bezrukov:2019mak}. One such mechanism, discussed in Ref. \cite{Berlin:2016bdv}, generates sterile neutrino DM driven by the dynamical evolution of an ultralight scalar. The large initial value of the scalar field generates a large mixing angle initially, enabling efficient production of sterile neutrinos with the correct abundance. However, the scalar field dilutes significantly during evolution, resulting in a tiny mixing angle today. Consequently, the sterile neutrinos are not detectable through X/$\gamma$-ray observations.

In this work, we highlight the potential of scalar-driven mechanisms to not only relax the X/$\gamma$-ray constraints from astrophysical observations but also to alleviate the constraints on WDM by generating a cooler spectrum. Specifically, we find that for large scalar mass, the scalar misalignment occurs earlier and dilutes its field value earlier, leading to the production of sterile neutrinos dominating at higher temperatures. We match the mass power spectrum of sterile neutrino DM with thermal relic WDM to obtain the corresponding constraints. The results show that the scalar-driven mechanism generates a much cooler spectrum for sterile neutrino DM compared to the DW mechanism, potentially helping to alleviate cosmological constraints on WDM. 
Moreover, it should be noted that in this model, the dominant form of DM is sterile neutrino, while the relic scalar produced by misalignment is negligible today.

Furthermore, in the original scalar-driven mechanism, the current mixing angle is typically too small to be probed by future X/$\gamma$-ray observations or other astrophysical searches. However, we discovered that introducing a new coupling between the sterile neutrino and ultralight DM can produce a density-dependent mixing angle today. 
The new term leads to a stationary value for the scalar field from the feedback of the existence of sterile neutrinos. This stationary term is larger than the small relic produced by misalignment, thus is the main contribution to the effective mixing angle today. Therefore, it changes the signal flux of the sterile neutrino decaying into SM neutrino and photon, making it depend on the DM density cubed rather than linearly, which helps evade the blank sky constraints and distinguishes it from normal decaying DM. Moreover, the new term provides a much larger flux for the X/$\gamma$-ray signal from sterile neutrino decay than the original term. Finally, we emphasize that another dominant decay channel of sterile neutrinos involves decaying into a scalar and Standard Model neutrino, which does not depend on the scalar field value but is instead controlled by the Yukawa coupling to the scalar. This channel produces a monochromatic neutrino signal, which is not covered by current neutrino telescope searches due to the small sterile neutrino DM mass, but it could be an interesting signal to crosscheck this model in the future, in addition to the X/$\gamma$-ray signal.

This paper is structured as follows. Section~\ref{sec:model} outlines the model setup, introducing the sterile neutrino DM and the scalar field together with their  interactions. In Section~\ref{sec:coupled-production}, we perform calculations for the coupled-evolution of these two components up to the present time. We then examine the constraints from various terrestrial and cosmological observations, focusing particularly on the comparison of the matter power spectrum between our model and thermal relic WDM, in Section~\ref{sec:constraints}. Next, in Section~\ref{sec:xray}, we investigate the X/$\gamma$-ray flux in our scenario and compare it with existing constraints. Finally, in Section~\ref{sec:conclusion}, we present our conclusions.

\section{Model setup} 
\label{sec:model}

To illustrate the impact of ultralight DM on the DW mechanism, we begin with a UV-complete model that includes a heavy doublet vector-like fermion, denoted by $\psi$, with SM gauge charge $(1,2,-1)$, a Majorana singlet sterile neutrino, represented by $N$, and an ultralight scalar, designated as $\phi$. The relevant UV-complete Lagrangian is as follows:
\begin{equation}
   -\mathcal L_\text{UV} \supset  \left[\tilde{y}_1\phi \bar{L}\psi+\tilde{y}_2\bar{\psi}\tilde{H}N^c  + \frac{1}{2} \lambda \phi \overline {N^c} N+h.c.\right] .
 \end{equation}
Integrating out the heavy fermion doublet $\psi$ produces a dimension-5 operator of the form $\tilde{y_1}\tilde{y_2}\frac{\phi}{\Lambda}\bar L \tilde{H} N^c +h.c.$. This operator, upon electroweak symmetry breaking, leads to the operator $y \phi \bar \nu N^c +h.c.$. The resulting low energy effective Lagrangian, with restored mass terms, can be expressed as follows: 
 \begin{equation}
     -\mathcal L = \left[ \frac{1}{2} (m_N + \lambda \phi)\overline{N^c} N + y \phi \bar \nu N^c +h.c. \right] + \frac{1}{2} m_\phi^2 \phi^2.
 \end{equation} 
 
Assuming that the scalar field $\phi$ is ultralight, it can be treated as a classical field that persists throughout the evolution of the universe. Taking this classical background field into consideration, we can diagonalize the mass terms and obtain the following result:
\begin{equation}
    \begin{aligned}\textbf{}
&-\mathcal{L}\supset \frac{1}{2} m_\phi^2 \phi^2 + \left( \frac{1}{2}\begin{pmatrix}
\overline{\nu^c} & \overline{N^c} 
\end{pmatrix}
\begin{pmatrix}
    0 & y \phi \\
    y \phi & \lambda \phi + m_N
\end{pmatrix}
\begin{pmatrix}
    \nu\\
    N
\end{pmatrix}+h.c. \right) \\
&\to\frac{1}{2} m_\phi^2 \phi^2 + \frac{1}{4}\left[\sqrt{(\lambda\phi+m_N)^2+4(y\phi)^2} -(\lambda\phi+m_N)\right]\overline{\nu^c}\nu \ \\ 
& + \frac{1}{4}\left[\sqrt{(\lambda\phi+m_N)^2+4(y\phi)^2} + (\lambda\phi+m_N)\right]\overline{N^c}N + h.c.,
\end{aligned}
\end{equation}
where we have performed a chiral rotation of the SM neutrino field to ensure that the mass term remains positive. We can represent the mass terms of $\nu$ and $N$ as follows:
\begin{align}
m_\nu (\phi) &= \frac{1}{2}\left[\sqrt{(\lambda\phi+m_N)^2+4(y\phi)^2} -(\lambda\phi+m_N)\right], \nonumber \\
m_N(\phi) &= \frac{1}{2}\left[\sqrt{(\lambda\phi+m_N)^2+4(y\phi)^2} + (\lambda\phi+m_N)\right].
\end{align}
After diagonalization, mixing between active and sterile neutrinos is induced and can be expressed as:
\begin{equation}
    \text{tan}\theta=\frac{y\phi}{m_N(\phi)}.
\end{equation}
This mixing is crucial in the production of sterile neutrinos in the early universe since the scattering between active neutrinos in the thermal bath can produce sterile neutrinos at a rate proportional to $\sin^2 \theta$.

\section{The production of sterile neutrino DM}
\label{sec:coupled-production}

\subsection{The Boltzmann equation for sterile neutrino}
The detailed description of the production process is given by the following Boltzmann equation, which is consistent with the equation in the original DW mechanism \cite{Dodelson:1993je,Abazajian:2005gj,Hansen:2017rxr,DeGouvea:2019wpf}:
\begin{align}
\nonumber
\frac{\partial}{\partial t}f_N(p,t)&-Hp\frac{\partial}{\partial p}f_N(p,t)
\approx\frac{1}{4}\Gamma_{\text{SM}}(p,T)\mathrm{sin}^2(2\theta_\text{eff})\left[f_\nu(p,t)-f_N(p,t)\right],
\label{Boltzmann-Eq}
\end{align}
with
\begin{align}
 \text{sin}^2 \left(2 \theta_{\text{eff}}\right)\equiv\frac{\Delta^2(p)\text{sin}^2 \left(2\theta\right)}{\Delta^2(p)\text{sin}^2(2\theta)+\Gamma_{\text{SM}}^2/4+\left(\Delta(p)\text{cos}(2\theta)-V^T(p)\right)^2}.
 \end{align}
where $f_{N/\nu}$ represents the momentum phase space distribution function, $H$ denotes the Hubble constant, and $\Delta(p)=(m_N^2-m_\nu^2)/2p\approx m_N^2/2p $ is the active neutrino oscillation factor in vacuum. The interaction rate for SM neutrinos can be parameterized $\Gamma_{\text{SM}}(p)= \lambda(p,T) G_F^2 p T^4$, where $G_F$ is the Fermi constant. We adopt the numerical results of $\lambda(p,T)$ given by Ref.~\cite{Asaka:2006nq} for electron neutrino.
Since there the sterile neutrino distribution compared with active one is negligible all over the production process, we therefore omit the feedback term. 
The neutrino thermal potential for flavor $\alpha$~\cite{Merle:2015vzu,Chu:2006ua,Abazajian:2001nj}
\begin{equation}
    V^T_\alpha(p)=-\frac{8\sqrt{2}G_F p_\nu}{3m_Z^2}\left(\langle E_{\nu^\alpha}\rangle n_{\nu^\alpha}+ \langle E_{\bar{\nu}^\alpha}\rangle n_{\bar{\nu}^\alpha}\right)-\frac{8\sqrt{2}G_F p_\nu}{3m_W^2}\left(\langle E_{\alpha}\rangle n_{\alpha}+ \langle E_{\bar{\alpha}}\rangle n_{\bar{\alpha}}\right)
\end{equation}
 is generated by SM electroweak interaction. The asymmetric lepton potential $V^L$, which arises due to the presence of lepton number asymmetry, can be expressed as follows:
\begin{equation}
    V^L=\sqrt{2}G_F\left[2(n_{\nu_\alpha}-n_{\bar{\nu}_\alpha} )+\sum_{\beta\neq\alpha}(n_{\nu_\beta}-n_{\bar{\nu}_\beta})-n_{\rm DM}/2\right].
\end{equation}
It is worth noting that in the presence of a non-negligible lepton number, the sterile neutrino could be produced through resonant production, as described in Refs.~\cite{ Shi:1998km, Abazajian:2005gj, Abazajian:2014gza}. In this study, we are restricted to the lepton number symmetric cases.

During the epoch of the QCD phase transition at a temperature around 170 MeV, as indicated by lattice simulations~\cite{Bernard:2004je, Karsch:2000kv}, the production rate of sterile neutrinos remains significant. Consequently, it becomes crucial to account for the sharp change in relativistic degrees of freedom in our calculations to accurately assess the impact on sterile neutrino production and spectra. The conventional method of treating $g_\star / g_{\star s}$ as constant is no longer valid in this scenario. To derive the general time-temperature relations, one can utilize the conservation of comoving entropy density, given by $g_{\star s}(T)T^3a(T)^3 = \textit{Const.}$, and differentiate this equation with respect to time, resulting in:
\begin{equation}
    \frac{dT}{dt}=-H(T)T\cdot\left(1+\frac{1}{3}\frac{d\ln g_{\star s}(T)}{d\ln T}\right)^{-1}.
    \label{time-temperature}
\end{equation}
By changing the variables from $f(t,p)$ to $f(T,p/(g_{\star s})^{1/3}T)$ and utilizing the time-temperature relations along with the identity that follows the trajectory $p/(g_{\star s})^{1/3}T = \textit{Const.}$, we obtain the following expression:
\begin{align}
    T\left(\frac{\partial f}{\partial T}\right)_p + p\left(1+\frac{1}{3}\frac{d\ln g_{\star s}(T)}{d\ln T}\right)\left(\frac{\partial f}{\partial p}\right)_T=T\left(\frac{\partial f}{\partial T}\right)_{p/(g_{\star s})^{1/3}T},
\end{align} 
After integrating the Boltzmann equation of sterile neutrinos over temperature, we can simplify Eq.(3.1) to the following form:
\begin{align}
   f_N(T_f,p)=\int_{T_\text{ini}}^{T_f}h\left(p\left(\frac{g_{\star s}(T_2)}{g_{\star s}(T_f)}\right)^{1/3}\frac{T_2}{T_f},T_2\right)dT_2.
   \label{boltzman-01}
\end{align}
with
\begin{align}
h(p,T)=\frac{1}{-4H(T)T}\Gamma_{\text{SM}}(p,T)\mathrm{sin}^2(2\theta_\text{eff})f_\nu(p,t)\cdot\left(1+\frac{1}{3}\frac{d\ln g_{\star s}(T)}{d\ln T}\right),
\end{align}
and $T_\text{ini}(T_f)$ represents the initial (final) temperature. This result agrees with Ref.~\cite{Merle:2015vzu}. The effects of the QCD phase transition are encoded in $\Gamma_\text{SM}(p,T)$ and $d\ln g_{\star s}(T)/d\ln T$.

\subsection{The evolution of the $\phi$ field}
The equation Eq.~\eqref{boltzman-01} describes the evolution of the sterile neutrino distribution in the early universe, which is strongly influenced by the evolution of the $\phi$ field. Therefore, it is necessary to establish the equation of motion for the $\phi$ field by 
\begin{align}
    \ddot{\phi}+3H\dot{\phi}+\frac{\partial V_\phi}{\partial \phi}=0, 
    \label{phi_evo}
\end{align}
with the potential $V_\phi$ given by
\begin{align}
    V_\phi=\frac{1}{2}m_\phi^2\phi^2&+ \frac{1}{2}m_N(\phi)\langle \overline{N^c} N + h.c.  \rangle -\frac{1}{\pi^2}T^4J_F\left[\frac{m_\nu(\phi)^2}{T^2}\right].
\end{align}
In the early universe, active neutrinos are thermalized and can contribute to the scalar potential through thermal loop effects~\cite{Batell:2021ofv}. However, due to the highly suppressed coupling between active neutrinos and $\phi$, the effect of this contribution is negligible. The expectation value of sterile neutrinos can also contribute to the scalar potential, even though they are out of equilibrium. When the sterile neutrinos cool down as the universe expands, we can take the non-relativistic limit $\langle \overline{N^c} N + h.c. \rangle \sim 2 n_{\rm DM}$. In this case,
\begin{align}
    \partial V_\phi/\partial \phi \simeq \lambda n_{\rm DM} +\left(m_\phi^2+\frac{2y^2n_{\rm DM}}{m_N}\right)\phi ,
\end{align}
is an excellent approximation for the parameters of interest. However, at earlier stages, when sterile neutrinos are still relativistic, a more careful consideration of the expectation value is required. It is convenient to rewrite the expectation value as $\langle \bar {N^c} N + h.c. \rangle = 2 \sigma(T) n_{\rm DM}$, and a good approximation is $\sigma(T) \sim 10^{-5}$ for $T \simeq 200 \text{MeV}$ when the most sterile neutrino are produced. 
However, given that $y$ must be very small based on the lifetime of DM via $N \to \phi + \nu$ decay (see Table~\ref{tab:parameters}), this contribution is also negligible in the production of DM. It is similar that one can neglect the tiny $\lambda$ term due to DM self-interaction constraint in Table~\ref{tab:parameters}. As a result, the $\phi$ field evolves as a free scalar field in the misalignment scenario.

\begin{table}[!tb]
    \centering
    \begin{tabular*}{0.9\textwidth}{@{\extracolsep{\fill}} ccccc}
    \hline\hline
    $\lambda$ & $y$  & $m_\phi$(eV)  & $\phi_\text{ini}$(GeV) &$m_N$(keV) \\
    \hline
    \\
    $10^{-24}$ & $\lesssim\mathcal{O}(10^{-18})$ & $\mathcal{O}(10^{-22}\sim10^{-9})$ & $\mathcal{O}(10^{8}\sim 10^{9})$&$\mathcal{O}(10\sim1000) $\\
    \\
    \hline\hline
    \end{tabular*}
    \caption{
    The parameters of this model are show in the table: $\lambda$ and $m_\phi$ are related to long-range interactions of DM, with the former subject to fifth force constraints, while $y$ is constrained by the sterile neutrino lifetime ($\tau_s>\tau_\text{Universe}$). The parameters $y$ and $\phi_\text{ini}$ are related to the dynamic mixing angle of the sterile neutrino, where the latter is critical for sterile neutrino production in the early Universe.    }
    \label{tab:parameters}
\end{table}

Consequently, the evolution of $\phi$ follows a mechanism that is similar to the misalignment of ultralight DM. Initially, it starts with a large value in the early universe, which remains constant due to the large Hubble friction. As the universe expands and cools, the field begins to oscillate around its minimum, with its amplitude decaying as $a^{-3}$. This evolution of $\phi$ leads to a significant mixing angle between sterile and active neutrinos in the early universe, resulting in the enhanced production of sterile neutrino DM while evading current constraints from the negligible mixing angle at the present time.

\begin{figure}[htb]
    \centering
    \includegraphics[width=0.325\textwidth]{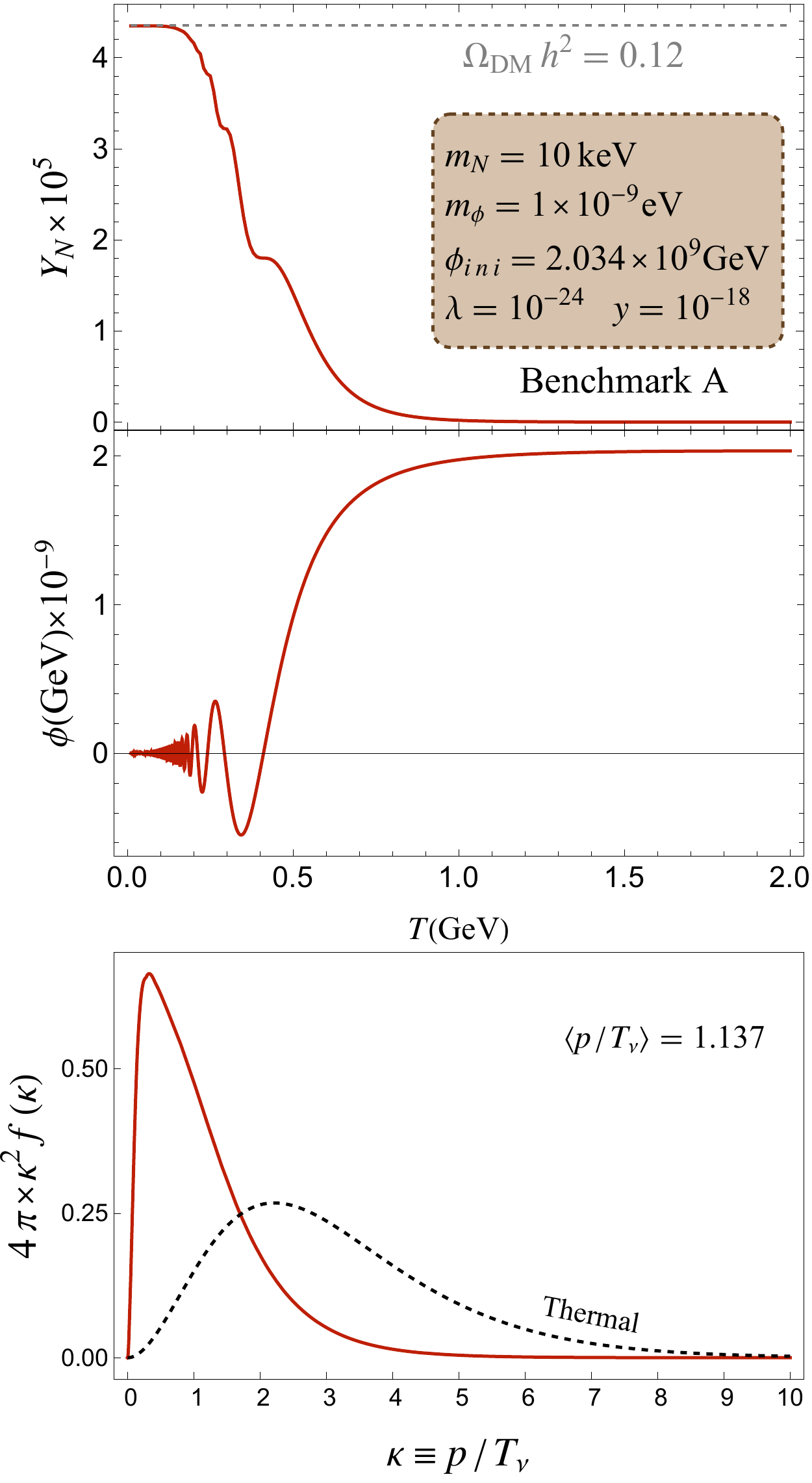}
    \includegraphics[width=0.325\textwidth]{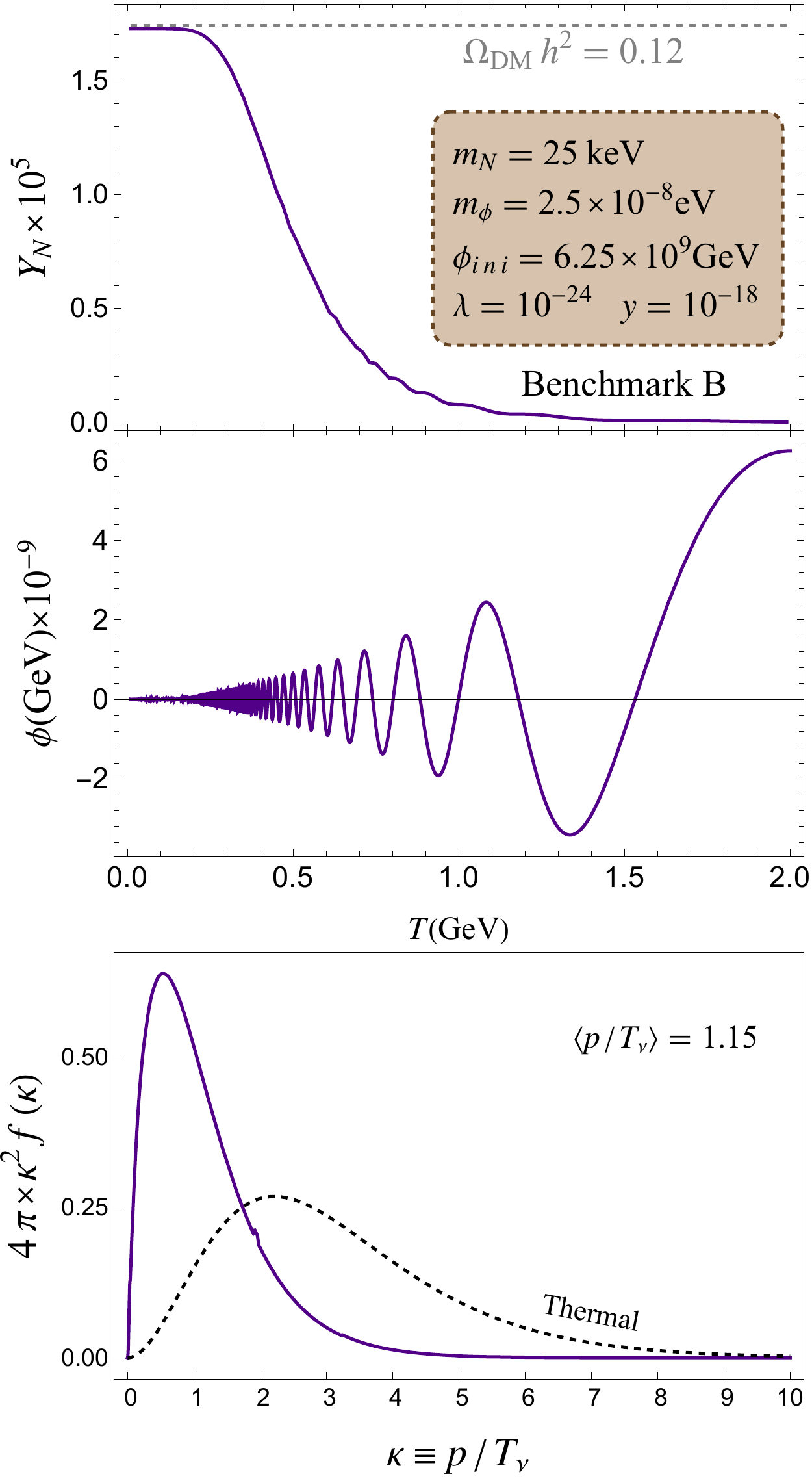}
    \includegraphics[width=0.325\textwidth]{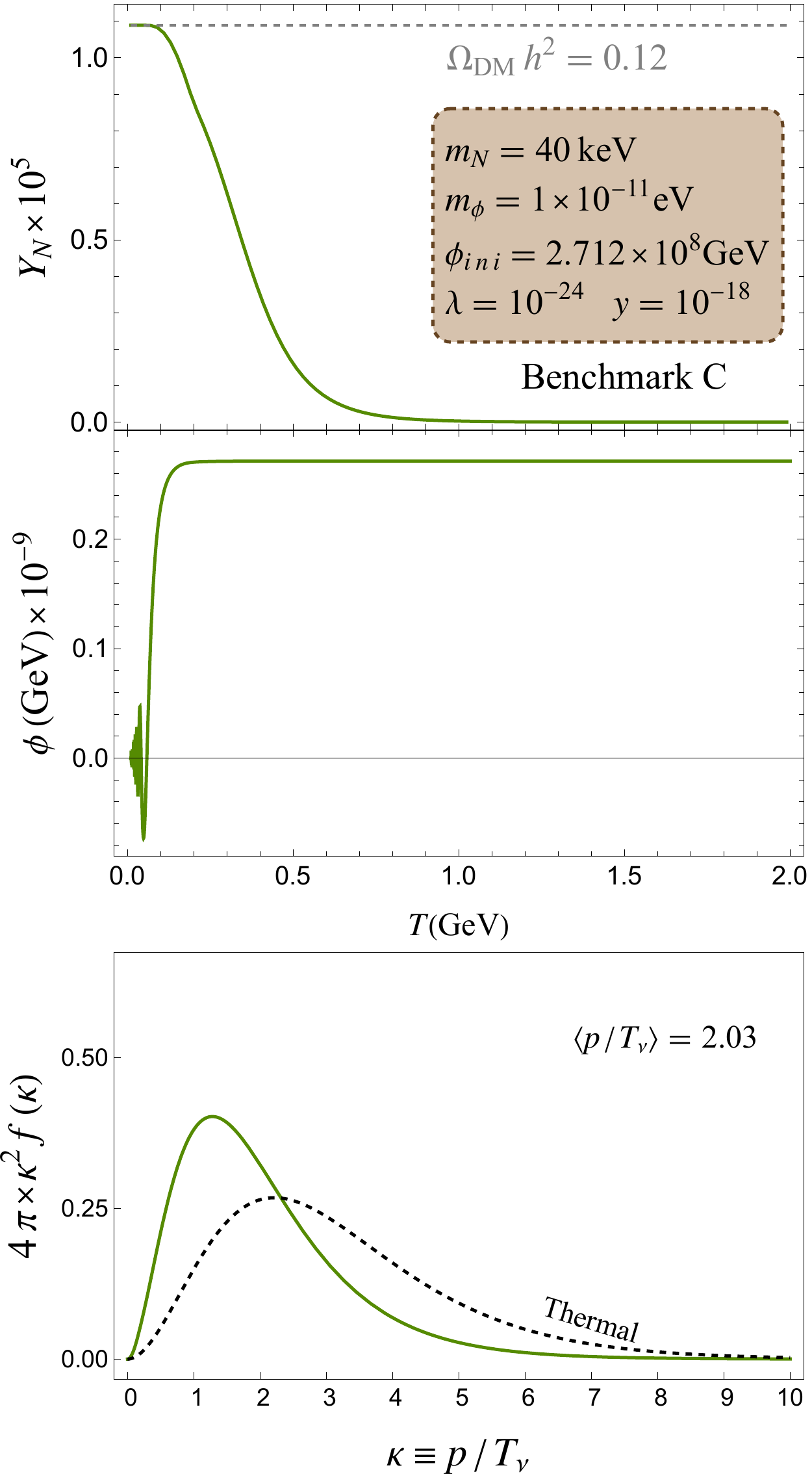}
    \caption{We present the evolution of  three explicit benchmarks (A, B and C) at early Universe, with the first two cases satisfy the condition $m_\phi >3H(T_\text{max})$. The $\phi$ field starts to oscillate earlier than the major production time of sterile neutrino, thus the evolution of $\phi$ deeply influenced the production of sterile neutrino. The upper panel shows the evolution of sterile neutrino number density, with yield $Y_N\equiv n_{N}/s$. The middle panel shows the evolution of the ultralight scalar $\phi$. The bottom panel shows the result phase space distributions for each benchmarks. We fix $\lambda = 10^{-24},~y=10^{-18}$ for all three benchmarks A (left), B (middle) and C (right) and choose parameters $\{m_{N}~(\text{keV}), ~m_\phi~(\text{eV}),~ \phi_\text{ini}~(\text{GeV}) \}$ equal to  $\{10,~1\times 10^{-9},~2.034\times10^9 \}$ for benchmark A, $\{ 25,~2.5\times 10^{-8},~6.25\times10^9 \}$ for benchmark B, and $\{40,~1\times 10^{-11},~2.712\times10^8 \}$ for benchmark C respectively.  These values enable the right DM relic abundance for $N$ and a negligible relic for $\phi$. Furthermore, the interference with $\phi$ and the change of entropy degree of freedom result in cooler spectrum compared with normalized  thermal Fermi-Dirac distribution which corresponding to DW sterile neutrino assuming constant $g_{\star s}$. 
        }
    \label{fig:phi-evolution}
\end{figure}

\subsection{The coupled evolution of sterile neutrino and $\phi$ field}

To take a more robust approach of evolution history, in principle one need solve the coupled differential equations of $\phi$ and $N$. 

Integrating Eq.~\eqref{boltzman-01} over momentum space and change the variable of Eq.~\eqref{phi_evo} to T using time-temperature relation Eq.~\eqref{time-temperature} we arrive at 
\begin{align}
&\phi''-\phi'\frac{g_{\star s}(T)'}{g_{\star s}(T)}+\phi\frac{(\partial V_\phi/\partial \phi)}{H(T)^2T^2}\cdot\left(1+\frac{1}{3}\frac{d\ln g_{\star s}(T)}{d\ln T}\right)^2=0, \\
&~~~~n_{\rm DM}(T_f)=\frac{1}{(2\pi)^3}\int d^3\vec{p} \cdot f_N\left[T_f,p,\sin^2(2\theta_\text{eff})(\phi)\right],
\end{align}
 where the primes denote differentiation with respect to the temperature. This later equation is coupled to the former one through the function of effective mixing angle $\sin^2(2\theta_\text{eff})$ which depends on the field value of $\phi$, while the number density of $N$ could also modify the effective potential $V(\phi)$. 
However, the two equations can be treated as decoupled since the feedback terms from $N$ on the evolution of the $\phi$ field are negligible. Hence, we can first numerically solve the second equation and then use the evolution of $\phi$ in the first equation to determine the sterile neutrino density. In Figure~\ref{fig:phi-evolution}, we demonstrate the evolution of $\phi$ and dark matter yield $Y_N\equiv n_{\rm DM}/s$ as functions of temperature for three representative parameter points. Initially, the production of sterile neutrinos is suppressed due to the high temperature, which leads to the suppression of $\sin \theta_{\text{eff}}$. At the same time, the large initial value of $\phi$, $\phi_{\rm ini}$, remains frozen due to the large Hubble friction. As the temperature drops, the production of sterile neutrinos increases, and the Hubble friction decreases, allowing $\phi$ to begin its oscillation. The production process can take two different forms depending on whether the oscillation of $\phi$ begins before or after the maximal production rate of sterile neutrinos, which is denoted by $T_\text{max}$. The former case is referred to as $m_{\phi} > 3H(T_\text{max})$, while the latter case is known as $m_{\phi} < 3H(T_\text{max})$. For standard DW case the production rate peaks at $T_\text{max,DW}\simeq 133~\text{MeV}\times (m_N/\text{keV})^{1/3}$~\cite{Dodelson:1993je} and sharply decreases when temperature is away from this value. In this model, the production rate at early universe is enhanced, resulting to $T_\text{max}\simeq 2~T_\text{max,DW}$~\cite{Berlin:2016bdv}. The first and second benchmarks illustrated in Fig~\ref{fig:phi-evolution} belong to the former case and are chosen to ensure the production of an appropriate abundance of sterile neutrino DM at around $T\approx 170\text{MeV}$. In addition, the relic abundance of $\phi$ today is negligible comparing with the sterile neutrino.
For the $m_{\phi} < 3H(T_\text{max})$ case, the $\phi$ field behaves as an over-damped oscillator  and frozen to its initial value during the most of sterile neutrino production period. Since   the $\phi$ influence of sterile neutrino production is small, the production mechanism is much more like traditional DW scenario in this case. 
\begin{figure}[htb]
    \centering
    \includegraphics[width=0.7\textwidth]{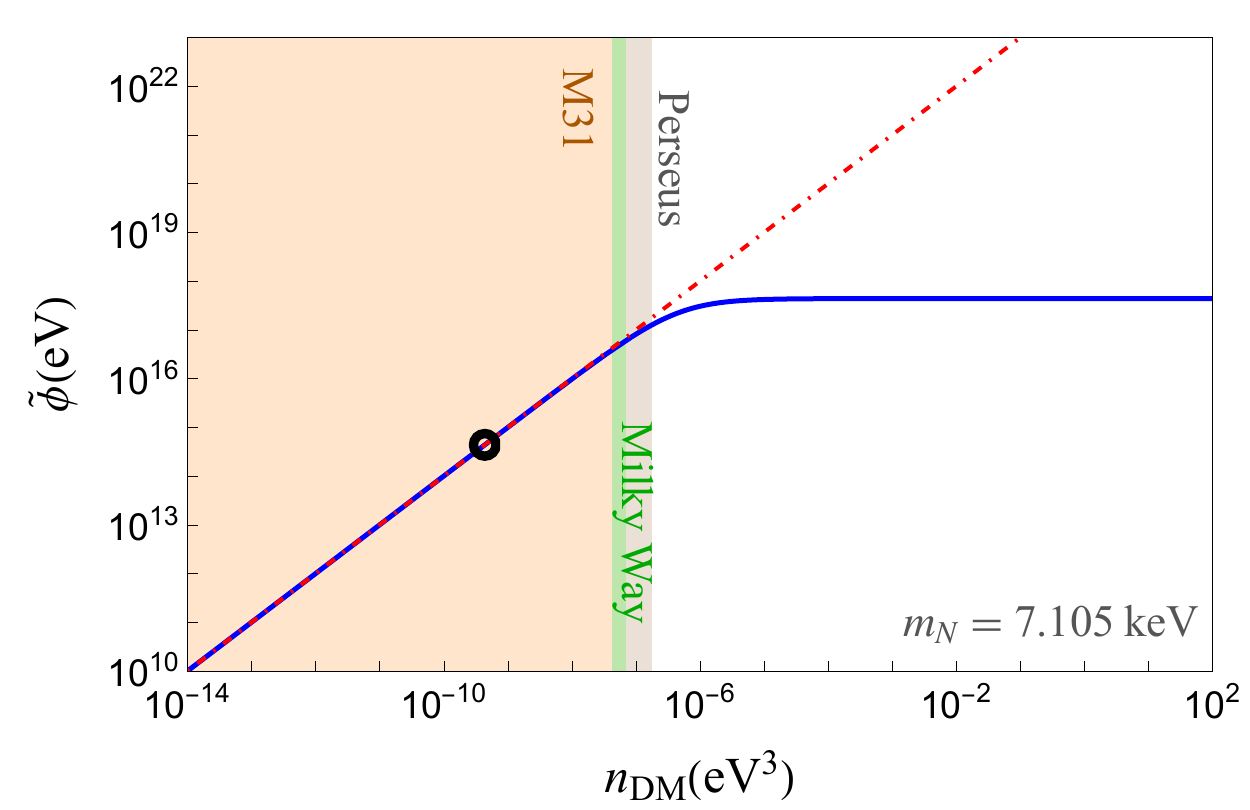}
    \caption{The stationary value $\tilde{\phi}$ as a function of DM number density $n_{\rm DM}$ in the present day. The red dot-dashed line represents the asymptotic behavior for small $n_{\rm DM}$. We set $m_\phi=10^{-24} \ \text{eV}, \ \lambda=10^{-24},$ and $y=9\times 10^{-20}$ for numeric demonstration. The colored regions correspond to different ranges of galactic DM number density, extending up to 0.1 kpc  under the assumption of the NFW profile~\cite{Navarro:1996gj}. The black circle denotes the stationary value at local dark matter density $\rho_{\rm DM}^{\rm local}=0.4~\text{GeV}/\text{cm}^3.$
    }
    \label{fig:vev}
\end{figure}
The energy density of the $\phi$ field today can be estimated by~\cite{Arias:2012az}
\begin{equation}
    \rho_\phi\simeq 1.7\times 10^{-16}\frac{\text{GeV}}{\text{cm}^3}\times \sqrt{\frac{m_\phi}{10^{-10}\text{eV}}}\left(\frac{\phi_\text{ini}}{10^{9}\text{GeV}}\right)^2\mathcal{F}(T_0),
    \label{rho_phi}
\end{equation}
where $\phi_\text{ini}$ is  the initial field value, $T_0$ is the temperature $\phi$ start to osccilate and $\mathcal{F}(T_0)\equiv (g_*(T_0)/3.36)^\frac{3}{4}(g_{*S}(T_0)/3.91)^{-1}$ is a smooth function of range $(0.3,1)$.
In the parameter space we are interested in, the contribution of energy density $\rho_\phi$ and its present day amplitude are very small. As a result, the ultralight scalar $\phi$ does not contribute to the DM relic abundance.

The present $\phi$ field, $\phi_0$, consists of two components in the following,
\begin{align}
   &~~~~~~~~~~~~~~~~~~~~~~~~ \phi_0\simeq \tilde{\phi}+\hat{\phi}~\mathrm{cos}(m_\phi t + \theta_0).\nonumber\\
    &\hat{\phi}=\sqrt{2\rho_\phi}/m_\phi,~~~~\tilde{\phi}=
    \begin{cases}
    \approx \lambda n_{\rm DM}/m_\phi^2 & n_{\rm DM}\ll n_{\rm DM}^{\rm max}\\
    \approx \frac{m_N(\sqrt{y^2(3\lambda^2+y^2)}-y^2)}{3\lambda y^2} & n_{\rm DM}\ge n_{\rm DM}^{\rm max}
    \end{cases}\\
    &~~~~~~~~~~~~~~~~n_{\rm DM}^{\rm max}\equiv \frac{2m_Nm_\phi^2(\sqrt{y^2(3\lambda^2+y^2)}-y^2)}{3\lambda^2 y^2}.
\end{align}

The first component, $\tilde{\phi}$, arises from the feedback of the local sterile neutrino and is generated through scalar-sterile neutrino interactions. The second component, $\hat{\phi}$, corresponds to the oscillation energy density produced from the misalignment, and $\theta_0$ denotes a random phase . The stationary value $\tilde{\phi}$ minimize the potential and thus depends on the number density of dark matter $n_{\rm DM}$ (see, \textit{e.g.} Ref.~\cite{Davoudiasl:2022ubh}). As shown in Fig.~\ref{fig:vev}, at low DM densities, the stationary value $\tilde{\phi}$ increases linearly with $n_{\rm DM}$, while at high densities $n_{\rm DM}\ge n_{\rm DM}^{\rm max}$, it saturates at a maximum value of $\tilde{\phi}_{\text{max}}$. This density-dependent behavior of $\tilde{\phi}$ has important implications for the decay rate of the sterile neutrino, preventing it from decaying too rapidly in the early Universe and in dense DM systems such as dwarf spheroidal galaxies (dSphs). Additionally, this density-dependent decay rate offers a promising signal for testing this scenario.

\section{Constraints from terrestrial and cosmological observations}
\label{sec:constraints}

The proposed model includes an ultralight scalar, $\phi$, that can mediate a long-range attractive force between sterile neutrino DM particles. This force could have observable effects on tests of the Equivalence Principle in the dark sector, using tidal tails~\cite{Kesden:2006vz,Kesden:2006zb}. It could also lead to astrophysical bounds~\cite{Davoudiasl:2018ltz,Morris:2013hua} that constrain the model parameter $\beta\equiv \lambda M_{\rm pl}/\sqrt{4\pi}m_N<2.2$, which is equivalent to $\lambda\lesssim 10^{-24}$ as shown in Table~\ref{tab:parameters}.

In addition, the Yukawa coupling $y$ leads to the sterile neutrino decay channel $N\to \nu + \phi$ with a decay width of $\Gamma(N\to\nu \phi)=y^2m_N/(16\pi)$. However, the power spectrum of the cosmic microwave background requires the DM lifetime to be longer than the age of the Universe~\cite{Poulin:2016nat}, thus constraining the parameter $y$ to be small $\lesssim 10^{-18}$ in Table~\ref{tab:parameters}. 
In addition, the decay $N\to \nu + \phi$ leads to a monochromatic neutrino flux. However, the active neutrinos with energies below MeV are less constrained in existing neutrino telescope data (see Ref.~\cite{Arguelles:2022nbl}), providing a parameter space for $m_N \lesssim 1~{\rm MeV}$.  This model could potentially be tested in future neutrino telescopes when searching for low energy neutrino flux.

Regarding neutrino self-interaction, it is mediated by $\phi$, whose strength is negligible in this model due to the small coupling $y$ and the mixing angle $\theta$. Therefore, it does not affect $\Delta N_{\text{eff}}$ and is consistent with CMB and BBN constraints. This is different from the model in Ref.~\cite{Kelly:2020aks}, which requires a modified effective thermal potential for neutrinos. Moreover, this model does not modify neutrino mass and oscillation in the parameter space considered, i.e. $\lambda \lesssim 10^{-24}$ and $y\sim \mathcal{O}(10^{-18})$, so limits from neutrino oscillation experiments do not apply.

\subsection{The colder energy spectrum of the sterile neutrino DM} 

Sterile neutrinos generated via the DW~\cite{Dodelson:1993je} mechanism typically exhibit a warm power spectrum that closely resembles that of their active counterparts. As WDM, they can significantly alter the matter power spectrum, resulting in the suppression of structures on small scales. In contrast to cold DM, sterile neutrinos produced through the DW mechanism exhibit unique velocity and phase space density distributions. Observations of small-scale structures have conclusively ruled out the existence of sterile neutrino dark matter with a mass below 92 keV.~\cite{Zelko:2022tgf}.

Various proposals have been put forward as potential solutions to the aforementioned issues. These include resonantly produced sterile neutrinos~\cite{Shi:1998km,Abazajian:2014gza}, decay from heavy particles~\cite{Abazajian:2019ejt}, and neutrino self-interaction~\cite{DeGouvea:2019wpf}. In our particular scenario, sterile neutrinos arise from an ultralight scalar-driven mechanism, resulting in a power spectrum that is naturally colder. The phase space distribution function of these sterile neutrinos, at a given SM neutrino temperature $T$, can be described using the Eq.~\eqref{boltzman-01}.

In bottom panel of Fig~\ref{fig:phi-evolution}, it can be noticed that the production of sterile neutrinos in our proposed model is reliant on the evolution of the ultralight scalar field $\phi$. Under appropriate parameters, this can result in a distribution that is colder than that produced through the DW mechanism. For instance, if $\phi$ remains at its initial large value, $\phi_\text{ini}$, due to Hubble friction in the early Universe but begins to oscillate early, the production of sterile neutrinos through neutrino oscillations is forced to reach its maximum at high temperatures. As a result, a colder spectrum is obtained compared to sterile neutrinos DM generated through the DW mechanism, as illustrated in Fig.~\ref{fig:phi-evolution}. This case happens for $m_\phi > 3H(T_\text{max})$ and is in better agreement with observations of cosmological structure formation. However, this feature of a colder sterile neutrino spectrum is nearly absent in the other case where $m_\phi < 3H(T_\text{max})$. In this case, during most of the neutrino oscillation production, the $\phi$ field remains constant due to the large Hubble friction, resulting in the similar outcome as the DW mechanism.

\begin{figure}[htbp]
        \centering
        \includegraphics[width=0.69\textwidth]{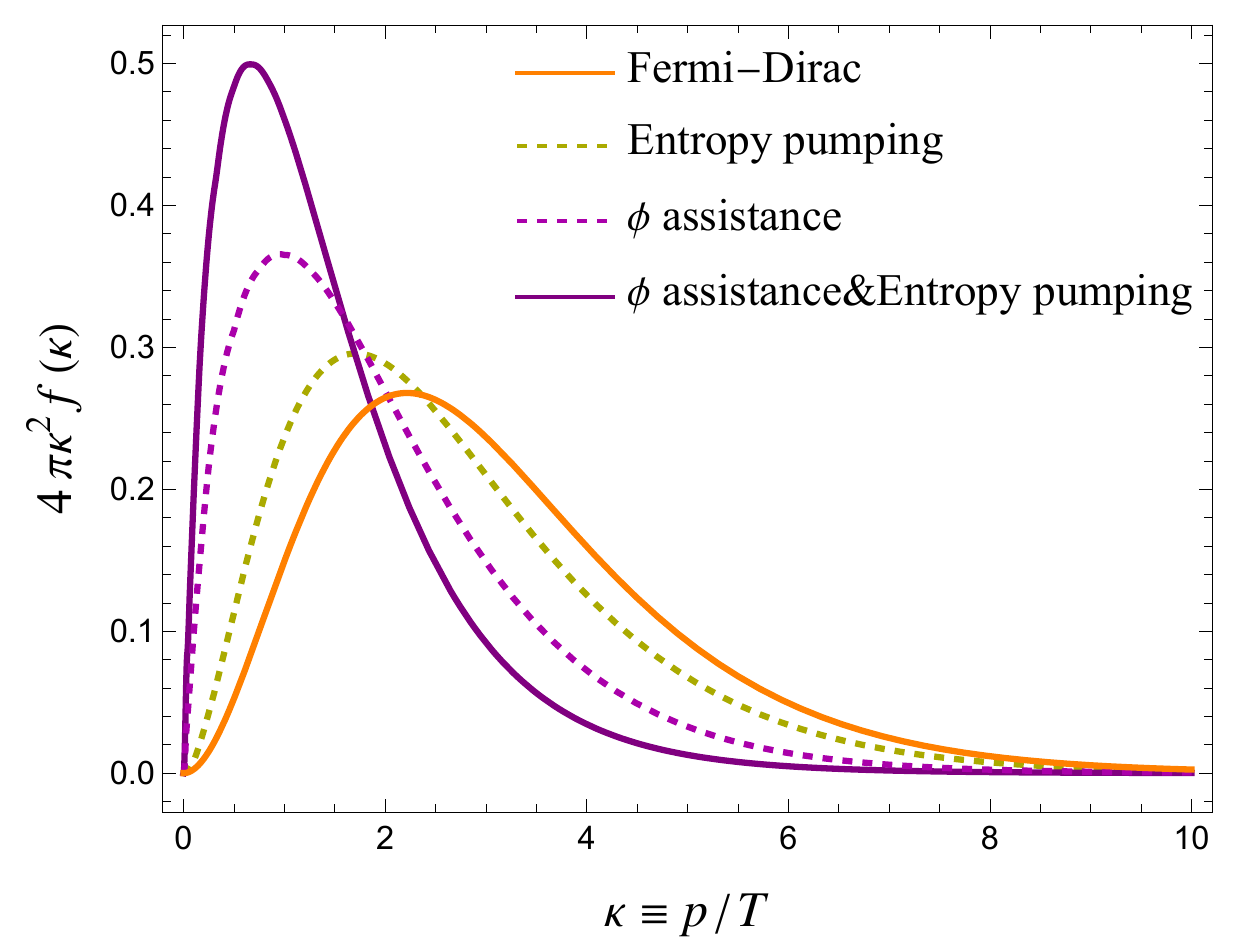}
    \caption{The influence of multiple factors on sterile neutrino spectrum with respect to thermal Fermi-Dirac case. We choose $m_N=10$ keV, $\sin^2(2\theta)=10^{-10}$ for demonstration. For $\phi$ assistance case, we set time dependent mixing angle $\sin^2 (2\theta)(T)=\sin^2(2\theta)_0 (T/T_0)^3$ using power law approximation, corresponding to $n=1$ case.
 }
        \label{fig:cold_factors}
\end{figure}

There are two primary reasons that lead to the cooler spectrum observed in our study. First, during the evolution of the universe, a change in degrees of freedom causes entropy pumping, effectively cooling the sterile neutrinos produced in the earlier stages. The second reason is related to the time-dependent evolution of the neutrino mixing angle and the specific features of the Dodelson-Widrow (DW) like production mechanism, which we qualitatively explain below. These factors are illustrated in Fig.~\ref{fig:cold_factors}.

In the original DW paper \cite{Dodelson:1993je} Eq.8, assuming constant $g_\star$ and neglecting certain minor terms in the effective mixing angle, the sterile neutrino spectrum is given by:
\begin{equation}
\frac{f_N(\kappa)}{f_\nu(\kappa)} \propto \sin^2(2\theta) \kappa \int_{0}^{x_{ini}} \frac{1}{(1+x^2 \kappa^2)^2} dx,
\end{equation}
where $x \propto T^3$, $\kappa=p/T$, $f_\nu(\kappa)$ represents the thermal distribution of active neutrinos ($1/(e^\kappa+1)$), and $\sin^2(2\theta)$ is constant in the original DW case. It is evident that when $x_{ini}$ tends to infinity, the sterile neutrino distribution converges to the thermal one
\begin{align}
    f_N(\kappa)\propto f_\nu(\kappa)\propto \frac{1}{1+e^\kappa}.
\end{align}

However, in our case, $\sin^2(2\theta)$ is no longer constant but varies with time after the misalignment occurs. For $m_\phi > 3H(T_\text{max})$, misalignment happens early, and the energy density of $\phi$ behaves as non-relativistic matter $\rho_\phi \propto a^{-3} \sim T^{3}$ with relation between the scale factor and temperature in radiation dominated Universe being used at last.  Therefore the subsequent evolution of the $\phi$ field can be approximated as $\phi \propto T^{3/2}$, or equivalently, $\sin^2(2\theta) \propto T^3 \propto x$. This results in the sterile neutrino spectrum:
\begin{equation}
\frac{f_N(\kappa)}{f_\nu(\kappa)} \propto \kappa \int_{0}^{x_{ini}} \frac{x}{(1+x^2 \kappa^2)^2} dx,
\end{equation}
which, when $x_{ini}$ tends to infinity, leads to a colder distribution:
\begin{equation}
f_N(\kappa) \propto \frac{1}{\kappa} f_\nu(\kappa) \propto \frac{\kappa^{-1}}{1+e^\kappa}.
\end{equation}
The singular behavior at $\kappa=0$ corresponds to the $x_{ini} \to \infty$ approximation, but the form of the distribution accurately characterizes the suppression of the spectrum at larger $\kappa$, thereby qualitatively explaining its cooling effect.

Moreover, we can further cool the spectrum by modifying the coupling between $\phi$ and neutrinos. If the mixing follows a general form of $\frac{\phi^n}{\Lambda^{n-1}}\bar {\nu} N^c$, then the $\phi$-dependence of the mixing angle changes to $\sin^2(2\theta) \propto \phi^{2n}$. In cases where misalignment occurs early enough, the scaling of the mixing angle becomes $\sin^2(2\theta) \propto T^{3n} \propto x^n$, resulting in the following sterile neutrino distribution:
\begin{equation}
f_N(\kappa) \propto \frac{1}{\kappa^n} f_\nu(\kappa) \propto \frac{\kappa^{-n}}{1+e^\kappa}.
\end{equation}

This flexibility allows us to achieve an even colder spectrum, further enriching our understanding of the role of neutrino mixing in shaping the final distribution.

To establish a connection between the constraints on small-scale structures from the thermal relic WDM model and the scalar-driven model under consideration, we employ the approach presented in Refs.~\cite{Zelko:2022tgf, Viel:2005qj, Abazajian:2019ejt, Abazajian:2005gj}. This approach relies on the transfer function, a crucial quantity that links the two scenarios. The transfer function describes the impact of free-streaming of warm dark matter on the matter power spectrum and is defined as follows:
\begin{equation}
\label{eq:transfer}
\hat{T}_{\text{WDM}}(k)\equiv \sqrt{\frac{P(k)}{P_{\text{CDM}}(k)}},
\end{equation}
where $P(k)$ and $P_{\text{CDM}}(k)$ represent the matter power spectrum of the specified WDM model and cold DM, respectively. The full Boltzmann code calculations of the linear power spectrum for thermal relic WDM(thWDM) have been performed in Refs.~\cite{Bode:2000gq, Viel:2005qj}. They reveal that the transfer function of thermal relic dark matter can be accurately approximated by the following analytic formula:
\begin{equation}
\label{eq:thWDM}
\hat{T}_{\text{thWDM}}(k) = [1+(\alpha k)^{2\mu}]^{-5/\mu},
\end{equation}
where $k$ is the comoving wavenumber, the shape parameter $\mu=1.12$, and $\alpha(m_{\text{thWDM}})$ is given by the expression:
\begin{equation}
\alpha(m_{\text{thWDM}})=0.049\left(\frac{m_{\text{thWDM}}}{\text{keV}}\right)^{-1.11}\left(\frac{\Omega_{\text{thWDM}}}{0.25}\right)^{0.11}\left(\frac{h}{0.7}\right)^{1.22}.
\end{equation}

WDM is characterized by a non-negligible velocity dispersion, which results in the suppression of the power spectrum below its characteristic free-streaming length. Consequently, the transfer functions of WDM exhibit a similar shape to that of cold DM at large scales, where WDM behaves more like a cold dark matter component. However, at specific small scales, an exponential cut-off occurs in the WDM transfer function due to the free-streaming process erasing perturbations. As a result, the transfer function of WDM shares a similar overall shape. It is important to acknowledge that the precise shape of WDM transfer functions depends on various details of the WDM model, such as its production mechanism and thermal history. These factors influence the location and sharpness of the cutoff in the transfer functions. Nonetheless, in general, the differences between warm and cold DM transfer functions are small~\cite{Vogel:2022odl}. 

Moreover, Ref.~\cite{Colombi:1995ze} showed that the transfer functions of DW sterile neutrino and thermal relic WDM are identical, which led Ref.~\cite{Viel:2005qj} to derive a one-to-one mass correspondence $m_{\rm DW}=4.43~\text{keV} (m_{\rm thWDM}/1~\text{keV})^{4/3}(0.25(0.7)^2/\Omega_{\rm DM}h^2)^{1/3}$. To accurately assess the impact of our model on the matter power spectrum and structure formation, we implemented the modified sterile neutrino energy distribution function in the non-CDM module of \texttt{CLASS}~\cite{Blas:2011rf} to compute the transfer functions.

\begin{figure}[htbp]
        \centering
        \includegraphics[width=0.49\textwidth]{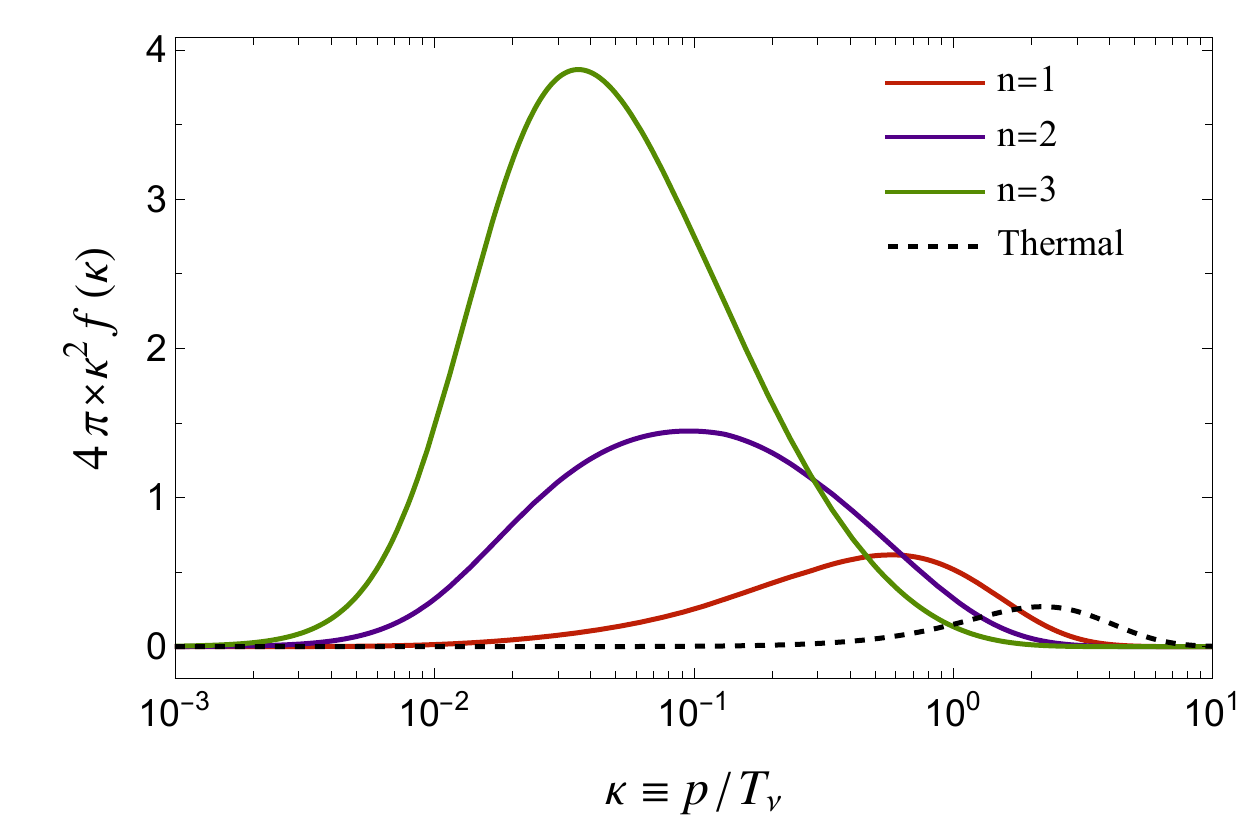}
        \includegraphics[width=0.49\textwidth]{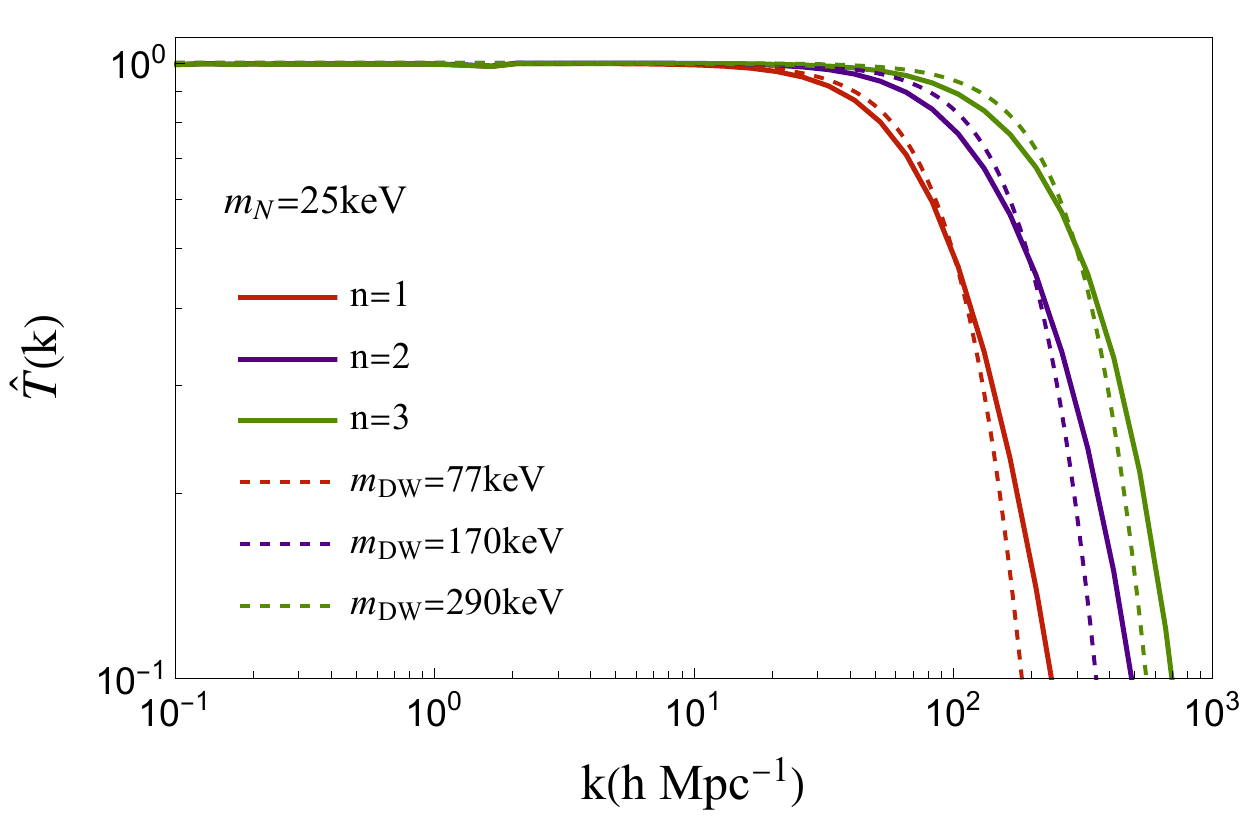}
    \caption{
\textit{Left panel}: The normalized phase space distribution with $\langle p/T_\nu \rangle \sim ~1.20,~0.60,~0.32$ for $n=1,~2,~3$ respectively. We choose $m_N= 25 ~\text{keV}$ for demonstration. \textit{Right panel}: The transfer functions of three choices of $n$ (represented by solid lines) are shown here, along with the transfer functions of three DW sterile neutrino particles (represented by dashed lines) at masses $m_{\rm DW} = 77,~170,$ and $290$ keV, respectively. Due to the similarity between the two sets of transfer functions, one can readily apply the small structure constraints on the masses of DW sterile neutrino particles to sterile neutrinos in this scenario.  }
        \label{fig:transfer}
\end{figure}

Fig.~\ref{fig:transfer} shows the energy spectrum and transfer function of sterile neutrino dark matter with mass $m_N=25 \  \text{keV}$ and different powers $n$ of $\phi$ in the $N-\nu$ mixing term. The figure reveals that a colder dark matter spectrum can be achieved for large enough $n$. We use the half-mode scale $k_\text{hm}$, defined as the wave number where $P(k)$ drops to $25\%$ of $P_\text{CDM}(k)$, to match our model with the thermal relic DM with the same $k_\text{hm}$ as in Ref.~\cite{An:2023mkf,Zelko:2022tgf}. The strongest limit on the thermal relic WDM mass excludes $m_{\text{thWDM}}<9.8 \ \text{keV}$, corresponding to a lower bound of $m_N > 92 \ \text{keV}$ for the original DW sterile neutrino DM. However, for our model with $n=2,3$, and $m_N=25 \ \text{keV}$, we obtain a spectrum equivalent to $m_{\text{thWDM}}=15.4,23 \ \text{keV}$, which satisfies the constraints. Table~ \ref{tab1} lists the corresponding masses of thWDM, DW sterile neutrino and three different values of $n$ in our scenario with $m_N=25  \ \text{keV}$. These results demonstrate that the scalar-driven scenario can produce a colder spectrum for sterile neutrino DM, which could potentially evade the small-scale structure constraints.

\begin{table}[tp]
    \centering
    \begin{tabular}{ c| c| c}
    \hline
         thWDM &DW sterile neutrino & Sterile neutrino in our model  \\
      \hline
      8.5 keV & 77 keV & 25 keV(n=1)   \\
      15.4 keV & 170 keV & 25 keV(n=2)   \\
      23 keV & 290 keV & 25 keV(n=3)  \\
         \hline
    \end{tabular}
     \caption{The correspondence among masses of this model and thWDM, DW sterile neutrino, through  matching transfer functions. The transfer functions of thWDM and DW sterile neutrino yield an one-to-one correspondence $m_{\rm DW}=4.43~\text{keV} (m_{\rm thWDM}/1~\text{keV})^{4/3}(0.25(0.7)^2/\Omega_{\rm DM}h^2)^{1/3}$~\cite{Viel:2005qj}.}
    \label{tab1}
\end{table}

\section{Constraints from X/$\gamma$-ray observations}
\label{sec:xray}

The decay of sterile neutrino DM into a photon and an active neutrino at the 1-loop level can produce a distinctive monochromatic photon signal that can be detected through astrophysical observations. This signal can be utilized to place constraints on the mixing angle between sterile and active neutrinos. The decay rate of a Majorana sterile neutrino is directly proportional to the square of the mixing angle, as described by the formula~\cite{Pal:1981rm},
\begin{align}
\Gamma_{N\to \nu \gamma} =\frac{9\alpha G_F^2}{2048\pi^4}\text{sin}^2 \left(2\theta \right)  m_N^5,
\end{align}
where we use $\theta$ instead of $\theta_{\rm eff}$ because the decay is occurring in the present-day rather than in the thermal environment of the early universe.

As DM particles are considered non-relativistic at present, the resulting energy of the photon in the final state is approximately half of the mass of the DM. The photon flux observed can be determined through integration along the line of sight:

\begin{equation}
F=\frac{\Gamma_{N\to \nu \gamma}  }{4\pi m_N} \int d\Omega_\text{f.o.v.}\int_\text{l.o.s}dr ~ \rho_\text{DM}\left(\sqrt{d^2+r^2-2dr\text{cos}\varphi}\right) ,
\end{equation}
where $d$ represents the distance of the source, and $\varphi$ denotes the angle between the line of sight and the galactic center. $\Omega_\text{f.o.v.}$ represents the field of view of the telescope. The relationship between $\varphi$ and the galactic coordinates $(l,b)$ is given by $\cos(\varphi)=\cos(l)\cdot \cos(b)$. The traditional DW sterile neutrino parameter space for making up $100\%$ dark matter relic abundance is completely ruled out by existing X-ray telescope observations~\cite{Abazajian:2017tcc}.

In our scenario, the decay rate of the monochromatic X/$\gamma$-ray signal $N\to \gamma\nu$ is proportional to the square of the mixing angle, $\mathrm{sin}^2(2\theta)$. However, the signal is integrated over the line of sight and the local mixing angle $\theta$ depends on field strength $\phi$, which has a different phase at each location. Thus, the time-average of it must be utilized to calculate the flux which is given as:
\begin{equation}
     \left \langle \mathrm{sin}^2(2\theta)\right \rangle \simeq 4\left \langle \frac{y^2 \phi_0^2}{m_N^2}  \right \rangle=\frac{4y^2}{m_N^2}\left(\frac{\lambda^2n_{\rm DM}^2}{m_\phi^4}  + \frac{\rho_\phi}{m_\phi^2}\right).
 \end{equation} 
 
In our study, we take into account the constraint arising from the fifth force between dark matter, the local dark matter density $\rho_\text{DM}^\text{local}=0.4 ~\text{GeV}/\text{cm}^3$ as a conservative estimation of $\rho_{\rm DM}$ in Galaxies and $\rho_\phi$ using Eq.\eqref{rho_phi}. Consequently, if the $m_\phi \ll \sqrt{\frac{\lambda^2 \rho_\text{DM}^2}{\rho_\phi m_N^2}}\sim 10^{-20}\ \text{eV}$, the former term on the right-hand side of $\left \langle \mathrm{sin}^2(2\theta)\right \rangle$ dominates, resulting in dark matter density-dependent mixing angles. This new characteristic offers a unique $X/\gamma$-ray signal. As the mixing angle is linearly dependent on the DM density in the parameter region of interest, the signal flux exhibits a novel feature $F \propto \int_\text{f.o.v.} \rho_{\text{DM}}^3 \ dV$. However, as demonstrated in Fig.\ref{fig:vev}, the mixing angle can saturate at large enough $n_{\rm DM}$, which implies that the estimate $F\propto \int_\text{f.o.v.} \rho_{\text{DM}}^3 \ dV$ may overestimate the flux, thus leading to more stringent constraints on the parameter space. Nevertheless, we will demonstrate later that despite some radical constraints, there is still a substantial parameter space available for this model. By substituting the formula for $\mathrm{sin}^2(2\theta)$, the flux in this model can be expressed as:

\begin{equation}
F=\frac{9\alpha G_F^2}{2048\pi^5}\cdot\frac{\lambda^2y^2}{m_\phi^4} \int d\Omega_\text{f.o.v.} \int_\text{l.o.s} dr ~ \rho^3_\text{DM}\left(\sqrt{d^2+r^2-2dr\text{cos}\varphi}\right) ,
\end{equation}
  As the flux is no longer dependent on the mass of the sterile neutrino, it suggests that X/$\gamma$-ray limits are less restrictive in higher sterile neutrino mass regions as compared to the normal decaying sterile neutrino scenario.

\begin{figure}
        \centering
        \includegraphics[width=0.9\textwidth]{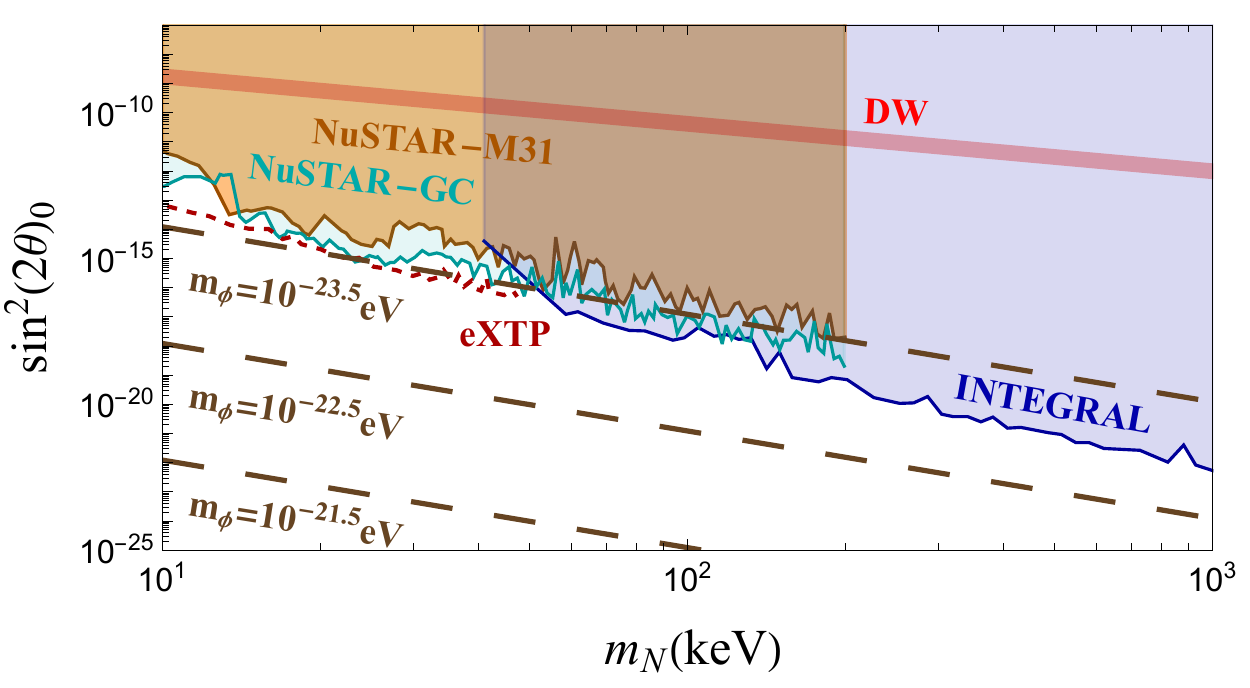}
    \caption{
    The parameter space of sterile neutrinos in this model. Since the mixing angle $\theta$ is dynamic and density dependent, we define $\mathrm{sin}^2(2\theta)_0\simeq 4(\lambda \  y \ \rho_\text{DM,local})^2/(m_\phi^2m_N^2)^2 $ with $\rho_\text{DM,local}=0.4~\text{GeV}/\text{cm}^3 $ around Earth for demonstration. Based on the mixing angle density dependent feature we adopt X-ray and Gamma ray observation constraints from NuSTAR Collaboration observations of Milky Center and M31 Andromeda Galaxy~\cite{Ng:2019gch,Perez:2016tcq} and null sterile neutrino search result from INTEGRAL~\cite{Boyarsky:2007ge}. The parameter space for original DW sterile neutrino making up all the dark matter is shown as a red band, which is completely ruled out by $\mathrm{X}$-ray observations~\cite{Abazajian:2017tcc}. In addition we plot several brown contours by fixing parameters $y,~\lambda$ to satisfy condition $\Gamma_s(y)\lesssim 1/\tau_\text{Universe}$ and $\lambda M_{Pl}/\sqrt{4\pi}m_N\lesssim 2.2$, in which the sterile neutrino in this model generate the observed relic density of DM. Some of the model parameter space can be probed by the proposed eXTP X-ray telescope Wide Field Monitor(WFM) observations of Galaxy blank sky region.~\cite{Malyshev:2020hcc} 
    }
        \label{fig:recast}
\end{figure}

To analyze the existing signals, we need to specify the DM profile. In this work, we adopt the DM profiles estimated by Ref. \cite{Cline:2014vsa}. To be conservative, we avoid the singularity at the center of the NFW profile by assuming that $\rho_{\text{DM}}(r<0.1 \text{kpc})=\rho_{\text{DM}}(0.1 \text{kpc})$. Additionally, we consider the contribution of the Milky Way DM flux within the telescope field of view for extra-galactic sources. The X-ray/$\gamma$-ray constraints are presented in Figure~\ref{fig:recast}, which includes the constraints from Refs.~\cite{Ng:2019gch,Perez:2016tcq,Boyarsky:2007ge}. 
The $y$-axis represents the sterile neutrino mixing angle $\sin^2(2\theta)_0$ defined at the solar system with a DM density of $\rho_{\text{DM}}^{\rm local}=0.4~\text{GeV}/\text{cm}^3$~\cite{deSalas:2020hbh}. We plot three lines of $\sin^2(2\theta)_0$ as a function of $m_N$ with three different choices of $m_\phi$. We ensure the correct DM relic density by choosing appropriate $\phi_{\rm ini}$, while fixing the parameters $\lambda$ and $y$ by maximizing the signal and satisfying the lifetime constraint $\Gamma_N(y)\lesssim 1/\tau_\text{Universe}$ and the fifth force constraint $\lambda M_{Pl}/\sqrt{4\pi}m_N\lesssim 2.2$ when $\lambda_\phi$ the de Broglie wavelength of $\phi$ is cosmological scale.

Future X-ray telescopes, such as Athena~\cite{Neronov:2015kca} and eXTP~\cite{Malyshev:2020hcc,Zhong:2020wre}, have the potential to explore some of the parameter spaces in our model. To illustrate this, we present the rescaled projected sensitivity of the eXTP telescope in Figure~\ref{fig:recast}. The eXTP X-ray telescope will be equipped with a Wide Field Monitor (WFM), which features a wide, steradian-scale Field of View (FoV) and operates in the $2-50$ keV energy range with an energy resolution of approximately $\sim 250$ eV and an effective area of about $\sim 80 ~ \text{cm}^2$. The WFM's extremely large FoV ($\sim 2.5 \ \text{sr}$) can significantly enhance decaying dark matter searches~\cite{Zhong:2020wre}. In this context, we rescale the projected search result using $1 \text{Msec}$ WFM blank sky observations from Ref.\cite{Malyshev:2020hcc}, and the corresponding results are shown as dashed red lines in Figure\ref{fig:recast}. With the advantages of a clean blank sky environment and a large FoV, the eXTP telescope is capable of probing some previously uncovered parameter spaces.

\section{Conclusion}
\label{sec:conclusion}

In this work, we provide a new version of the scalar-driven sterile neutrino production, where the oscillation production of sterile neutrino is modified by the interaction with the dynamical evolving ultralight scalar field in the early universe. We demonstrate that this approach can effectively suppress the production of sterile neutrinos at low temperatures with a heavy scalar mass, leading to a colder matter power spectrum that avoids constraints from small-scale structure observations. Notably, the dominant DM relic in this model is from the sterile neutrino, with only a small fraction coming from the ultralight scalar.
Furthermore, the model predicts a larger X/$\gamma$-ray flux with a light scalar mass, which flux is directly proportional to the cubic density of local sterile neutrinos. This feature distinguishes our model from normal decaying DM and enables it to avoid the blank-sky type limits. In addition, this model predicts a potential monochromatic neutrino signal that can complement the X-ray/$\gamma$-ray signal and may be detected by future neutrino telescopes.

\section*{Acknowledgement}
We would like to thank Yue Zhang for the helpful discussions. The work of J.L. is supported by the National Science Foundation of China under Grants No. 12075005 and No. 12235001, and by Peking University under startup Grant No. 7101502458. The work of X.P.W. is supported by the National Science Foundation of China under Grant No. 12005009 and the Fundamental Research Funds for the Central Universities.

\bibliographystyle{utphys}
\bibliography{ref}
\end{document}